\documentclass[usenatbib, usegraphicx]{mn2e} 
\usepackage{hyperref}

\begin{document}

\title{Predicting Coronal Mass Ejections transit times to Earth with neural network}
\author[D. Sudar et al.]{D. Sudar
        B. Vr\v{s}nak
        M. Dumbovi\'{c}
\\
Hvar Observatory, Faculty of Geodesy,
              Ka\v{c}i\'{c}eva 26, University of Zagreb, 10000 Zagreb, Croatia\\
}

   \date{Release \today}
   \maketitle

\begin{abstract}
Predicting transit times of Coronal Mass Ejections (CMEs) from their initial parameters
is a very important subject, not only from the scientific perspective, but also because
CMEs represent a hazard for human technology. We used a neural network to analyse transit times
for 153 events with only two input parameters: initial velocity of the CME, $v$,
and Central Meridian Distance, CMD, of its associated flare. We found that transit
time dependence on $v$
is showing a typical drag-like pattern in the solar wind. The results show that the speed
at which acceleration by drag changes to deceleration
is $v\approx$500 km s$^{-1}$. Transit times are also found to be
shorter for CMEs associated with flares on the western hemisphere than
those originating on the eastern side of the Sun. We attribute this difference
to the eastward deflection of CMEs on their path to 1 AU.
The average error of the NN prediction in comparison to observations is $\approx$12
hours which is comparable to other studies on the same subject.

\end{abstract}

\section{Introduction}
\label{SectIntro}
Coronal mass ejections (CMEs) are important drivers of space weather. They can cause strong
geomagnetic storms \citep{Gosling1990,
Gosling1991, Zhang2003, Echer2008, Richardson2012, Cid2014}.
\citet{Zhang2007} concluded that 87\% of strong geomagnetic storms were
caused by either a single CME or multiple interacting CMEs, while
\citet{Richardson2001} attribute 97\% of the most intense storms
to transient structures associated with CMEs.
Another significant cause of geomagnetic storms are corotating interactive regions
(CIRs) (see \citealt{Alves2006} and references therein).
CME-associated events can cause serious damage to communication and
navigation satellites, threaten the safety of astronauts,
and in most extreme cases can disrupt electric power supply on the ground
\citep{Boteler1998, Schrijver2013}.
Solar energetic particle (SEP) 
events accelerated by CME-driven shocks also present a risk to equipment and
humans in space (see \citealt{Dierckxsens2015} and references therein).
A more detailed summary of space weather
hazards is given in \citet{Feynman2000}.


Predicting when CMEs will reach the Earth is thus very important.
Typical transit times ($TT$) for CMEs to reach the Earth are between 1 and 5 days
\citep{Richardson2010}. In this paper, we use "CME" to refer to CMEs both near the Sun and
interplanetary space (often called "interplanetary" coronal mass ejections (ICMEs)
in other studies). Furthermore, observed transit times are based on the arrival of
the leading edge of the CME/ICME.
Understanding how the $TT$ depends on initial properties of the
CME and the associated phenomena is the most promising way of getting an
early warning for a possible geomagnetic storm. Unsurprisingly, velocity
of the CME in the LASCO field of view was found to be one of the most
significant parameters. 
\citet{Schwenn2005} derived an empirical logarithmic expression for $TT$ as a
function of the expansion velocity of CME for 75 events.
The standard deviation of the fit was about 14 hours.
\citet{Gopalswamy2001} derived a
semi-empirical model of $TT$s based on an effective interplanetary acceleration
and CME initial speed.
\citet{Fry2003} compared the $TT$ predicted by three different theoretical
models and found that the average error of all models is between
11 and 12 hours.

\citet{Vrsnak2004} have studied the kinematics of more than 5000 CMEs
between 2 and 30 solar radii. They have shown that even close to the
Sun the motion of CMEs is affected by aerodynamic drag through the
interaction with the ambient plasma. Drag based model \citep{Vrsnak2007, Vrsnak2013}
was a logical extension of this idea into interplanetary space where
the drag force is even more dominant as gravity and Lorentz force
weaken further away from the Sun.
Recently, \citet{Vrsnak2014} compared $TT$ predictions from the
Drag Based Model (DBM) and WSA-ENLIL+Cone Model \citep{Odstrcil2004, Taktakishvili2009}.
The mean value of the difference between calculated $TT$s for the
two models was $\overline{\Delta}=0.09\pm9$ hours, while
the average absolute difference from actual observations was $\approx$14
hours for both models.
\citet{Mays2015} used ensemble modelling to predict arrival time of CMEs.
For the 17 events which did reach the Earth, they found the average absolute error
to be $\approx$ 12.3 hours.
For a more detailed review of CME $TT$ predictions see \citet{Zhao2014}.

Another important effect of the ICME propagation through the heliosphere is
a deflection from the radial trajectory. \citet{Gosling1987} studied 19
fast CMEs and found a small eastward deflection of about 3\degr.
\citet{Wang2002} found that ~60\% of geoeffective CMEs occurred on the
western solar hemisphere. They attributed this asymmetry to the deflection
of CMEs in the interplanetary space.
\citet{Wang2004} concluded
that CMEs are affected by the Parker's spiral magnetic field and
deflected from the radial trajectory. In their model, fast CMEs are
deflected toward east, while slow ones are deflected toward west.
By studying latitudinal deflection (North-South direction) of CMEs,
\citet{Shen2011} and \citet{Gui2011} developed the magnetic energy density gradient
(MEDG) model. In their model, CMEs are deflected to the regions with
lower magnetic energy density.
Analysing driverless shocks, \citet{Gopalswamy2009} suggested that
CMEs are deflected from their radial trajectory by coronal holes (CH).
Based on the numerical analysis of the 2005 August 22 event, \citet{Lugaz2011}
concluded that the CME was deflected by the CH.

In this work we will use a neural network (NN) to analyse CME $TT$s as a function
of CME initial speed and central meridian distance (CMD). 
Neural networks have many applications and have been successfully used
in the analyses of astrophysical problems. For example, \citet{Prsa2008} used NN to
determine physical properties of eclipsing binaries, \citet{Li2013} used it for
solar flare forecasting, \citet{Valach2009} for quantifying the geomagnetic response
to particular solar events while \citet{Uwamahoro2012} used NN to
estimate the geoeffectiveness of halo CMEs.
The most advantageous aspect of using NNs for fitting observed data is that
it is not necessary to specify any functional form of the empirical curve.

\section{Data}
We compiled a list of 153 CMEs for which their ICME counterparts were detected and
their $TT$ to Earth was measured.
We used
only the events for which the CME source position could be
determined.
All CME-ICME pairings were taken from the catalogue provided by \citet{Richardson2010}.
For the arrival time of the ICME
we used times based on observations of plasma and field.
To determine the source position of the CME we consulted a number of papers
\citep{Gopalswamy2001, Zhang2003, Manoharan2006, Zhang2007, Marubashi2015} and we
also used the automatic method described in \citet{Vrsnak2005}.

\begin{table*}
\caption{Full list of events used in this paper is given as online material.}
\label{Tab_List}
\begin{tabular}{llccc}
\hline
Start time & Arrival Time & $TT$ [h] & $v$ [km s$^{-1}$]
& Source position\\
\hline
1996 Dec 19 1515 & 1996 Dec 23 1700 & 97.74 & 469 & S14W09\\
1997 Jan 6 1246 & 1997 Jan 10 0400 & 87.23 & 136 & S18E06\\
1997 Feb 7 0017 & 1997 Feb 10 0200 & 73.71 & 490 & S20W04\\
1997 Apr 7 1400 & 1997 Apr 11 0600 & 87.98 & 878 & S30E19\\
1997 May 12 0405 & 1997 May 15 0900 & 76.91 & 464 & N21W08\\
...\\
\hline
\end{tabular}
\end{table*}
As one model parameter we used the CME plane of the sky
speed, $v$, which was taken
from LASCO coronograph observations for each event\footnote[1]{\url{http://cdaw.gsfc.nasa.gov/CME_list/}}.
For the positional information about the onset of the CME
we used only CMD from the associated flare data
in order to investigate if there is an average effect of CMD on $TT$.
We assigned negative value of CMD to flares
on the eastern hemisphere and positive CMD to those appearing on the western hemisphere of
the Sun.
For the onset time of the CME we used the linear extrapolation with previously determined
speed back to the position of the associated flare. $TT$ is then given by the difference between
the arrival time determined by \citet{Richardson2010} and this onset time.
The full event list is included in the supplementary on-line materials; the first few
entries are shown in Table~\ref{Tab_List}.

\begin{figure}
\resizebox{\hsize}{!}{\includegraphics{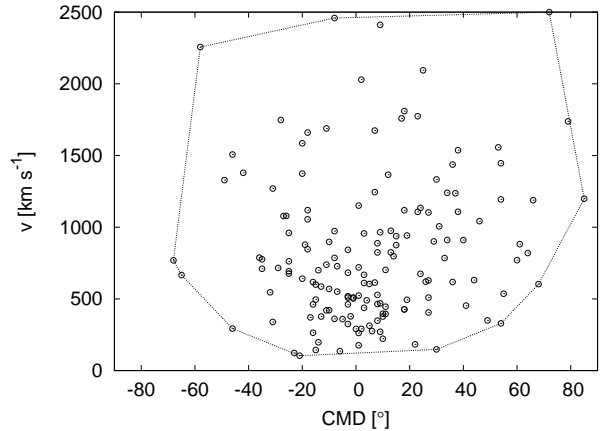}}
\caption{Distribution of velocities and CMDs in parameter space is shown with open circles,
while the dotted line marks the smallest convex shape that bounds all the events.}
\label{Fig_v_cmd_distr}
\end{figure}
In Fig.~\ref{Fig_v_cmd_distr} we show the distribution of CMEs in our sample
in the parameter space spanned by values of their initial velocity, $v$, and CMD.
The dotted line in the graph marks the smallest convex shape
which includes all the events we used in this work.

\section{Neural network method}
In this section we give a short overview of NNs and its application
to our problem.
For a more thorough discussion about NN refer to \citet{Gurney1997}
and references therein.

\begin{figure}
\resizebox{\hsize}{!}{\includegraphics{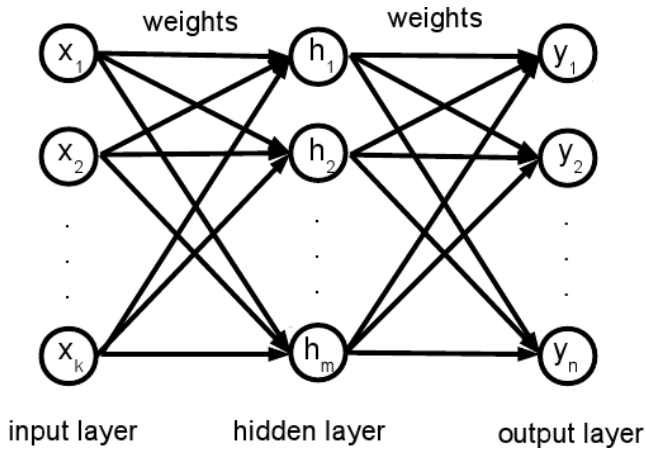}}
\caption{Schematic diagram of a typical feed forward neural network.}
\label{NN_scheme}
\end{figure}

We used a multilayer NN with feed-forward algorithm to transform
input to output parameters.
Schematic diagram of such
a network is shown in Fig.~\ref{NN_scheme}. NN takes $k$ input values, $x_{i}$, and
through a series of connections transforms the input into one or more output
values, $y_{i}$. Between input and output layer there can be one or more hidden layers,
although more than two are almost never needed. In Fig.~\ref{NN_scheme} we show
the weights as arrows connecting cells (denoted as circles) between different layers.

To calculate the value in $j$th hidden cell, $h_{j}$, from given input values $x_{i}$ we used
the following relation:
\begin{equation}
h_{j} = f\left(\sum\limits_{i=1}^k w_{ij}x_{i}\right),
\label{Eq_hidLayerCalc}
\end{equation}
where $w_{ij}$ are weights connecting cells between different layers and $f$ is one of the sigmoid functions (discussed below).
Values in the cells of the output layer
are calculated in the similar fashion using values in the cells of the hidden layer:
\begin{equation}
y_{j} = f\left(\sum\limits_{i=1}^m w_{ij}h_{i}\right).
\label{Eq_outLayerCalc}
\end{equation}
For the sigmoid function, we have chosen the logistic function:
\begin{equation}
f(x) = \frac{1}{1 + e^{-x}}.
\label{Eq_logistic}
\end{equation}
Sigmoid function has an 'S' shape with values bound by asymptotes at $y=0$ and $y=1$ for $x$ between
-$\infty$ and $\infty$, respectively.
Without the non-linear sigmoid function, the result of the NN would just be
an elaborate linear combination of input parameters.
Apart from introducing nonlinearity into the method, it also bounds the output
from all cells, so in practice input and output parameters are scaled by simple transformations
to cover the range of the sigmoid function used.
NN performance is not noticeably affected by the choice of the
sigmoid function. 

As we can see, weights, $w_{ij}$, are previously unknown and represent free parameters
of the NN model. If we use past events, known as samples in NN terminology, with measured input and output values we
can use backpropagation to optimise the weights. Backpropagation is a form of steepest
descent method which slowly optimises weights starting from random values.

Number of weights quickly grows with the number of cells used. By choosing a very large number
of cells and, consequently weights, it is possible to create a NN which will perfectly
fit all the available data. However, this is clearly not a statistically meaningful
procedure. Therefore, the data are divided in at least two samples: learning sample
and validation sample. Backpropagation is performed only on the learning sample while the
error of the validation sample reveals if we have encountered the overfitting problem.
Overfitting is caused when NN describes random noise rather than the underlying
relationship. This is typically revealed when the error of the validation sample starts
to grow with the increasing number of weights. It is also desirable that average errors
of the learning and validation samples have similar values.

From our 153 data points we have grouped
130 of them into the learning sample and the rest into the validation sample.
We have also used bias cells in the input and hidden layer. Bias cells are special
cells which connect only to forward layers and have a value of one. They are
useful for shifting the values of the sigmoid function and in general improve the
convergence of the fit.
The number of weights, $N_{w}$, is given by:
\begin{equation}
N_{w} = (N_{inp} + 1)N_{hid} + (N_{hid} + 1)N_{out},
\end{equation}
where $N_{inp}$, $N_{hid}$ and $N_{out}$ are the number of cells in
input, hidden and output layers, respectively. Bias cells are represented
by +1 in both brackets.

Statistical rule of the thumb to pick adequate number of weights is to have the ratio
between number of data points and number of weights, $N_{ld}/N_{w}\geq 5$ at least.
Of course, the higher the ratio, the results would be more statistically reliable.
Ratio of $\approx$10 or higher is recommended.

We used $v$ and CMD as input parameters, while the output
parameter was $TT$.
By choosing two input cells and
one output cell we can see that we are very limited in choosing the number
of cells in the hidden layer. We decided to use three cells in the hidden layer
which gives an acceptable ratio of $N_{ld}/N_{w}=130/13=10$.

\section{Results}
\subsection{Convergence and errors}
\begin{figure}
\resizebox{\hsize}{!}{\includegraphics{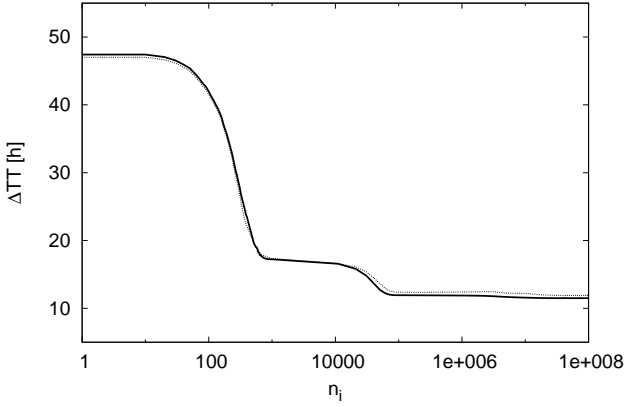}}
\caption{Improvement of the average error, $\Delta TT$, with each iteration
is shown for the learning sample and the validation sample with
a solid and dotted line, respectively.}
\label{Fig_errorsConverge}
\end{figure}

In the first step NN assigns random values to weights which typically results in
very large differences between calculated and observed output values. Then it
uses backpropagation on the learning sample to gradually improve the weights.
In Fig.~\ref{Fig_errorsConverge} we show the average errors of the learning
and the validation sample with increasing number of iterations.
We can see that the improvement occurs very quickly at the beginning of the
iteration process and after that average errors of both samples stabilise.
There is no exact rule when to stop the iterations, so we decided to end
the process when errors remain fairly constant for large number of iterations.

We ran this procedure ten times for different learning and validation samples, meaning
that samples were made randomly from the full sample of 153 events.
The average errors of the learning sample was varying between 10.89 and 11.99 hours,
while the average error of the validation sample were between 9.75 and 16.28 hours.
This means that, no matter which events are in the learning sample and from
which initial random weights the NN starts, the end result is always similar and stable.
It also means that our choice of the number of hidden cells was statistically
sound.

\begin{figure}
\resizebox{\hsize}{!}{\includegraphics{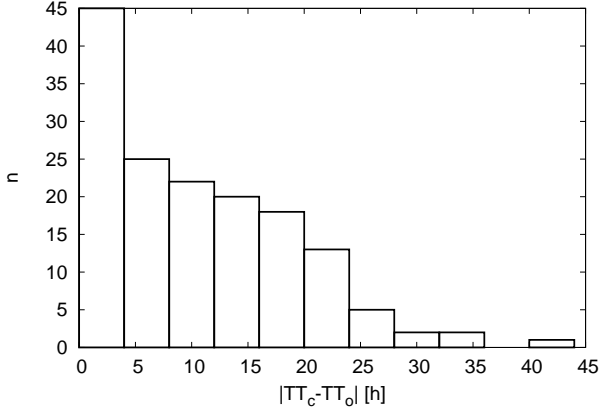}}
\caption{The distribution of absolute difference between
calculated and observed transit times, $|TT_{c}-TT_{o}|$,
for the whole sample.}
\label{Fig_absDistr}
\end{figure}

In Fig.~\ref{Fig_absDistr} we show distribution of absolute difference between
calculated and observed transit times, $|TT_{c}-TT_{o}|$,
for the learning sample. The bins are 4 hours
wide and we can see that most of the events
are contained in the first three bins ($\leq$12 hours).

From the ten different runs of NN optimisation we picked the one
with the lowest error of the whole sample as the best fit.
\begin{figure}
\resizebox{\hsize}{!}{\includegraphics{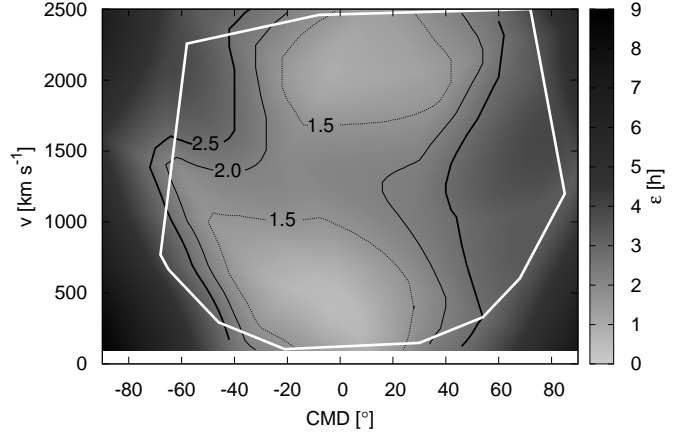}}
\caption{Map plot of error, $\varepsilon(v, CMD)$, based
on the best fit NN curve and other nine NN curves (Eq.~\ref{Eq_errorNN}).
Thick solid, thin solid and dotted isolines show the contours for $\varepsilon =2.5$ h,
$\varepsilon =2.0$ h and $\varepsilon = 1.5$ h, respectively. White solid line marks
the smallest convex shape bounding all the events.}
\label{Fig_error_map_plot}
\end{figure}
We can estimate the reliability of the fit by calculating the
absolute average differences between best fit values and nine
other values found by other NN runs for the full range of
input parameters:
\begin{equation}
\label{Eq_errorNN}
\varepsilon(v_{m}, CMD) = \frac{1}{9}\sum_{i>1}^{10}|TT - TT_{i}|.
\end{equation}
In Fig.~\ref{Fig_error_map_plot} we show a
map plot of $\varepsilon(v_{m}, CMD)$ as well as the smallest convex
polygon which encloses all the events with the white solid line
(see Fig.~\ref{Fig_v_cmd_distr}).
We can see that errors, $\varepsilon(v, CMD)$, are the smallest
within the enclosing polygon. Outside of this polygon errors can
become fairly large.
This illustrates the problem of extrapolating
the results with NN. Extrapolation is almost never recommended
with NN and its predictions outside of the input domain
are known to be very unreliable.

\subsection{Transit times prediction by NN}

\begin{figure}
\resizebox{\hsize}{!}{\includegraphics{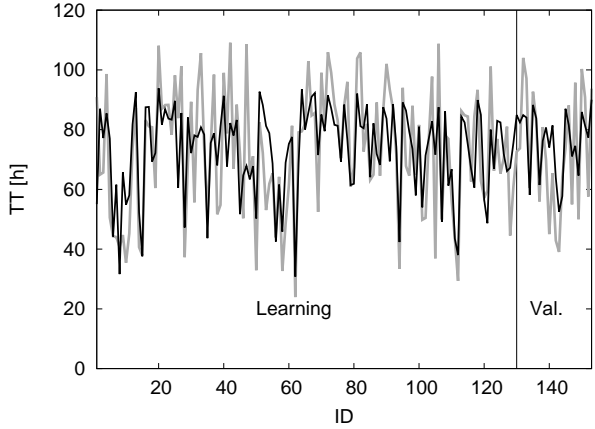}}
\caption{Observed $TT$ (grey line) is shown versus identification
number for learning and validation samples. Predicted $TT$ calculated
by NN are shown with a black line. Vertical line at
130 delimits the learning from the validation sample.}
\label{Fig_calc_obs_vsID}
\end{figure}

The most important feature of NNs is the agreement of its calculated
values and the actual observations.
In Fig.~\ref{Fig_calc_obs_vsID} we show calculated and observed
$TT$ versus event ID for both the learning and the validation
samples.
Observed $TT$ is shown with a thick solid grey line, while $TT$ calculated
by NN is shown with a solid black line line.
Observed $TT$s as a function of event ID, of course, looks like noise,
but we can see that it is closely matched by $TT$s predicted by our NN.

In the ideal case calculated transit times, $TT_{c}$, as a function
of the observed transit times, $TT_{o}$, should be a straight line
with the functional form $TT_{c}=TT_{0}$.
\begin{figure}
\resizebox{\hsize}{!}{\includegraphics{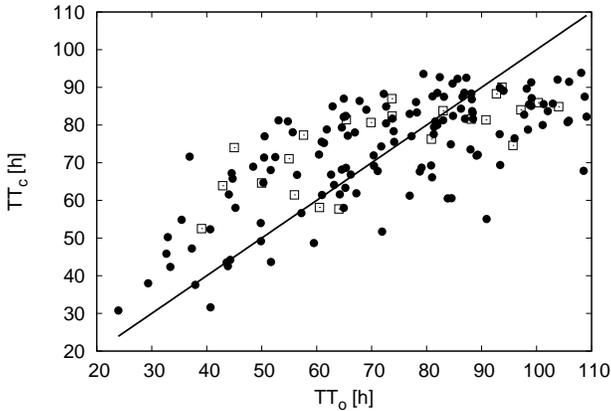}}
\caption{Calculated, $TT_{c}$, vs observed transit time, $T_{o}$.
Function $TT_{c}=TT_{0}$ is shown with the solid line. Filled circles and empty
squares are $TT$s from the learning and the validation sample, respectively.}
\label{Fig_TTsimVSreal}
\end{figure}
In Fig.~\ref{Fig_TTsimVSreal} we show this relationship. Filled circles
and empty squares represent $TT$ values from the learning and the
validation samples, respectively.
Function, $TT_{c}=TT_{0}$, is shown with a solid line.
The agreement looks satisfactory, but we can notice that there are no
$TT_{c}$ values below $\approx$30 hours and above $\approx$100 hours.
Also $TT_{c}$ are overestimated for $TT_{0}$ smaller than $\approx$ 80 hours.

\begin{figure}
\resizebox{\hsize}{!}{\includegraphics{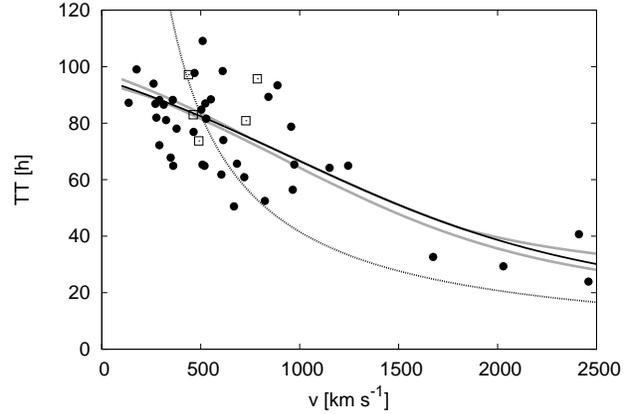}}
\caption{Transit times, $TT$, for CMEs near the central meridian. Filled circles
and empty squares
are observed $TT$s for CMEs in the region $-10\degr <$ CMD $<+10\degr$ in the
learning sample and the validation sample, respectively.
Solid black line represents $TT$s for CMD$=0\degr$ found by the best fit NN and solid grey lines
are maximum and minimum values of $TT$s found by all 10 NN fits.
The dotted line is a 'constant speed' $TT$ function.}
\label{Fig_DT_vs_vel_m10_p10}
\end{figure}

In Fig.~\ref{Fig_DT_vs_vel_m10_p10} we show the NN best fit curve for 
$TT$ versus CME speed, $v$, for central meridian with a black solid line.
With filled circles and empty squares we show observed $TT$s in the region $-10\degr <$ CMD $<+10\degr$
for the learning and validation samples, respectively. Grey solid lines represent maximum and minimum
values of all 10 NN fits as a function of initial CME velocity, $v$.
The difference between these two extreme values is larger where there are fewer
data points which is expected for NNs, as for any other fitting technique.

We can immediately see why the average error shown in Fig.~\ref{Fig_errorsConverge}
is fairly large ($\approx$ 12 h).
For a smaller error the curve would have to follow every measured $TT$ more closely.
However, such a curve would be unjustified from a statistical point of view and
as a result the error of the validation sample would be significantly larger
than the error of the learning sample. This occurs because NN can find a pattern
in otherwise random variations between different events and is known as overfitting.

In the same graph we also show a 'constant speed' model transit time, $TT=d/v$, with a
dotted line, where we have taken distance from the Sun to the Earth to be $d=149.6\cdot$10$^{6}$
km. Comparing this line with the one obtained with NN, we can see that CMEs slower
than $v\approx$500 km s$^{-1}$ arrive sooner than predicted by the 'constant speed' model.
Converse is true for CMEs faster than $v\approx$500 km s$^{-1}$.

\begin{figure}
\resizebox{\hsize}{!}{\includegraphics{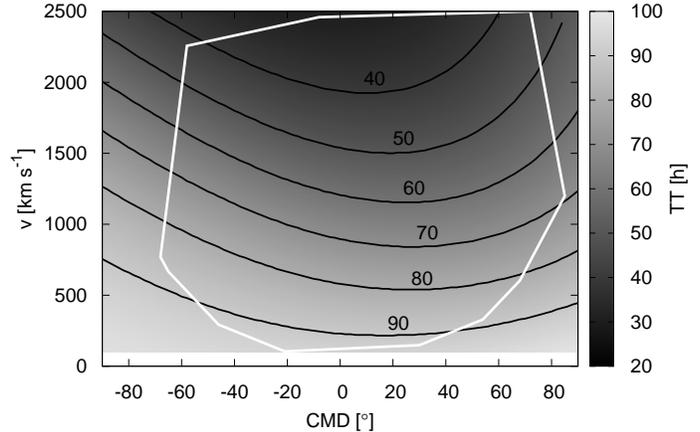}}
\caption{Map plot of $TT$ as function of $v$ and CMD. Solid lines
are $TT$ isolines with values of $TT$ in hours for each isoline shown
next to it. White solid line marks the smallest convex shape bounding
all the events in our data set.}
\label{Fig_TT_map_plot}
\end{figure}
In Fig.~\ref{Fig_TT_map_plot} we show the calculated $TT$s from the best fit as a function
of both input parameters, $v$ and CMD. Transit time isloines are shown as solid lines
with values of $TT$ in hours indicated next to each isoline.
Transit time, $TT$, is smaller for CMEs with larger velocities which was
expected, but we can also see that $TT$s are smaller for positive CMDs (western solar hemisphere)
in comparison with $TT$s for negative (eastern) CMDs.

\begin{figure}
\resizebox{\hsize}{!}{\includegraphics{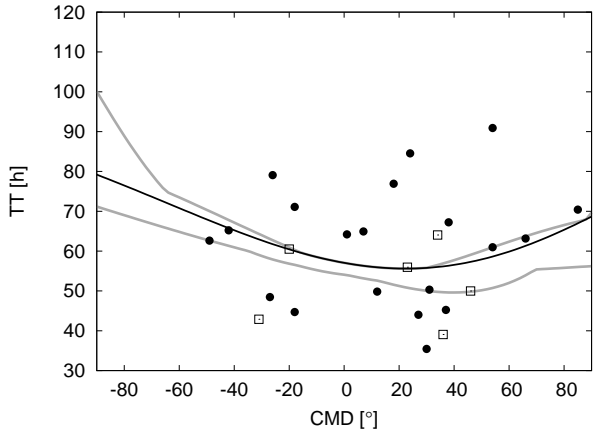}}
\caption{Transit times for CMEs with velocities between 1000 and 1500 km s$^{-1}$. Filled circles
and empty squares are observed $TT$s for CMEs in the region 1000 km s$^{-1} < L <$ 1500 km s$^{-1}$ from the
learning sample and the validation sample, respectively.
Solid black line are $TT$s for $v=1300$ km s$^{-1}$ found by the best fit NN and solid grey lines
are maximum and minimum values of $TT$s found by all 10 NN fits.}
\label{Fig_DT_vs_cmd_1000_1500}
\end{figure}
In order to verify the dependence of $TT$ on CMD we take a closer look of this phenomenon
in Fig.~\ref{Fig_DT_vs_cmd_1000_1500}. We used the range between 1000 km s$^{-1}$
and 1500 km s$^{-1}$ for data points because this bin has a good CMD coverage and still
a lot of data points (see Fig.~\ref{Fig_v_cmd_distr}).
As in Fig.~\ref{Fig_DT_vs_vel_m10_p10}, solid
circles are data points from the learning sample, while empty squares are events
from the validation sample. Black solid line is the result of the best fit NN for $v=1300$ km s$^{-1}$ and grey
lines are maximum and minimum values of all 10 NN fits as a function of CMD for the same speed.
The distribution of data points in Fig.~\ref{Fig_DT_vs_cmd_1000_1500}, assuming that it is not just an artefact of
small number of events, is shifted westward which also illustrates the east-west asymmetry already observed in Fig.~\ref{Fig_TT_map_plot}.
As before, we can see the difference between maximum and minimum values growing very quickly
in the regions with few or no data points.

\begin{figure}
\resizebox{\hsize}{!}{\includegraphics{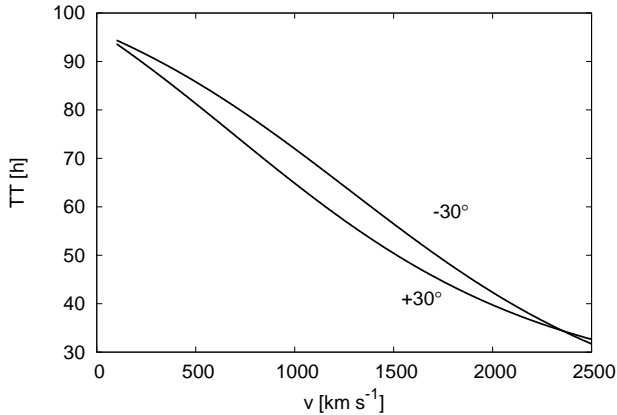}}
\caption{Best fit curves found by NN for CMD$=+30\degr$ and CMD$=-30\degr$.
The values of CMD parameters are given next to each curve.}
\label{Fig_DT_vs_vel_m30_p30}
\end{figure}
In Fig.~\ref{Fig_DT_vs_vel_m30_p30} we show the difference in calculated
$TT$ as a function of CME initial speed, $v$, for CMD$=+30\degr$ and CMD$=-30\degr$.
Calculated $TT$s for CMEs originating from the eastern hemisphere (negative CMD)
are larger for almost all speeds, $v$, than $TT$s that NN predicts for their western
counterparts.


\section{Discussion and Conclusion}

Arguably, the largest benefit of using 
NNs is that it is not necessary
to specify the empirical function or hyper-surface that relates the input and output
parameters. This becomes more difficult to determine as additional input parameters
are included.
Neural networks
do not have such problems as long as proper statistical reasoning is used.

Here we have applied a NN to estimating ICME transit times
using the LASCO CME speed and CMD source position from 153 events as input parameters.
The average error of the whole sample for the best fit is 11.56 hours,
which is comparable to other works (see Sect.~\ref{SectIntro} or \citet{Zhao2014}).
Additional error of $\approx$1 hour can be expected due to the rounding
of ICME arrival time to the nearest hour in most of the events in our dataset
\citep{Richardson2010}.
Looking at Figs.~\ref{Fig_DT_vs_vel_m10_p10} and \ref{Fig_DT_vs_cmd_1000_1500}
it is clear that average errors can not be much smaller.
Better results could be achieved with more input parameters to the NN. One parameter
which almost certainly plays a significant part is the variable solar wind speed
during CME transit through interplanetary space.
We also know that energy released in the associated flare plays a role in early dynamics of
CMEs \citep{Vrsnak2005} and it is quite possible that this effect is reflected
in their actual $TT$ to Earth. In our case the
number of data points at our disposal practically prohibits this approach.
It is also possible to improve $TT$ prediction by tracking CMEs to larger
distances, but this results in shorter lead times \citep{Mostl2014}.

The NN shown in this work only predicts when particular CME will reach 1 AU
and it can't be used to predict $TT$ to any other distance from the Sun.
Therefore, it is only usable as a predictor for objects close to 1 AU, for
example the Earth or STEREO satellites.
It also does not predict whether the CME will actually hit the Earth or how
geoeffective it will be.

The accuracy of CME TT forecasts is not significantly
improved using the NN method discussed here. There is no way that one
limited set of initial parameters can fit all the observed data much better
than with $\approx$12 hour average error \citep{Zhao2014} at the present state of
data quality and quantity.
Case studies can offer better accuracy with a richer data set and by taking into account
additional factors such as CME interactions
\citep{Lugaz2012,Temmer2012,Maricic2014} or
variable solar wind \citep{Temmer2011}, but they
usually depend on rare configurations
of spacecrafts and the results of such analyses are usually too late to make
any sort of real time predictions \citep{Zhao2014}.

The results of NN shown in Fig.~\ref{Fig_DT_vs_vel_m10_p10} are quite
convincing that the CME is subjected to the drag force in the moving
medium on its path through the interplanetary space. \citet{Vrsnak2007} already
proposed a simple aerodynamic drag model for the CME dynamics.
Moreover, in their Fig. 5, \citet{Vrsnak2007} show theoretical $TT$ curves
which are very similar to our empirical one in Fig.~\ref{Fig_DT_vs_vel_m10_p10}.
From the equation for the drag acceleration \citep{Vrsnak2007}:
\begin{equation}
a = -\gamma (v_{CME} - w)|v_{CME} - w|,
\label{Eq_drag}
\end{equation}
where $w$ is the speed of the moving ambient medium we can easily interpret
the intersect between the 'constant speed' model, $a=0$, and the NN curve
shown in Fig.~\ref{Fig_DT_vs_vel_m10_p10}. The intersect is actually
the empirical value of the solar wind speed obtained by NN which
for our analysis we can write with asymmetric error $w = 505^{+7}_{-8}$ km s$^{-1}$.

Figs.~\ref{Fig_TT_map_plot},~\ref{Fig_DT_vs_cmd_1000_1500} and~\ref{Fig_DT_vs_vel_m30_p30} show
significant dependence of $TT$ on CMD. On average, CMEs originating on
the western hemisphere (positive CMD) arrive sooner then their eastern
hemisphere (negative CMD) counterparts. From Figs.~\ref{Fig_DT_vs_cmd_1000_1500}
and~\ref{Fig_DT_vs_vel_m30_p30} we can estimate that the difference is up to $\approx$10 hours.
This result can be interpreted as a consequence of east-west deflection of CMEs.
Another clue that deflection of CMEs is really happening can be found in
our Fig.~\ref{Fig_v_cmd_distr}. The bounding surface containing all the
events as well as their distribution in the parameter space is asymmetric and shifted westward. This means
that CMEs on the eastern limb miss the Earth, while those on the western limb
still have a chance to hit it.
The same can be seen in the data published by \citet{Wang2004} who also
noticed this westward shift of the Earth-bounding CME source regions.

From the isolines in Fig.~\ref{Fig_TT_map_plot} it is
possible to infer how the CMD corresponding to the minimum $TT$ varies as a
function of CME speed, $v$.  While western CMDs were found for almost all speeds in the
NN training runs made for this study, the locations were not stable such that even a
qualitative assessment of the dependence of the CMD offset on $v$ is not advisable.
The problem of the exact location of the $TT$ minimum is also visible
in Fig.~\ref{Fig_DT_vs_cmd_1000_1500}.

In a case study of a fast CME, \citet{Rollett2014} concluded
that the CME evolves asymmetrically in exactly the same direction (eastward)
as we can see in our results.
In contrast to \citet{Wang2004}, which propose that slow and fast CMEs are
deflected in different direction, we see eastward deflection at all speeds.
It is also not clear how the CH effect on CME trajectory \citep{Gopalswamy2009,Lugaz2011}
could account for any sort of statistical asymmetry in the east-west direction.
\citet{Shen2011} proposed a model for latitudinal deflection of CMEs toward the solar
equator, but it is not clear if their model could explain average longitudinal
deflection that we see in our data.

\section*{acknowledgements}
This work has been supported by the Croatian Science Foundation under the project
6212 "Solar and Stellar Variability". CME catalogue used in this work
is generated and maintained at the CDAW Data Center by NASA and The Catholic University of
America in cooperation with the Naval Research Laboratory. SOHO is a project of international
cooperation between ESA and NASA.
We are also thankful to the anonymous referee for many useful comments and suggestions
which improved the quality of this paper.

\bibliographystyle{mn2e}
\bibliography{mn-jour,NN_CME} 

\begin{thebibliography}{46}
\expandafter\ifx\csname natexlab\endcsname\relax\def\natexlab#1{#1}\fi

\bibitem[{{Alves}, {Echer} \& {Gonzalez}(2006){Alves}, {Echer}, \&
  {Gonzalez}}]{Alves2006}
{Alves} M.~V., {Echer} E., {Gonzalez} W.~D., 2006, J. Geophys. Res. (Space
  Physics), 111, 7

\bibitem[{{Boteler}, {Pirjola} \& {Nevanlinna}(1998){Boteler}, {Pirjola}, \&
  {Nevanlinna}}]{Boteler1998}
{Boteler} D.~H., {Pirjola} R.~J., {Nevanlinna} H., 1998, Advances in Space
  Research, 22, 17

\bibitem[{{Cid} {et~al}\mbox{.}(2014){Cid}, {Palacios}, {Saiz}, {Guerrero}, \&
  {Cerrato}}]{Cid2014}
{Cid} C., {Palacios} J., {Saiz} E., {Guerrero} A., {Cerrato} Y., 2014, Journal
  of Space Weather and Space Climate, 4, A28

\bibitem[{{Dierckxsens} {et~al}\mbox{.}(2015){Dierckxsens}, {Tziotziou},
  {Dalla}, {Patsou}, {Marsh}, {Crosby}, {Malandraki}, \&
  {Tsiropoula}}]{Dierckxsens2015}
{Dierckxsens} M., {Tziotziou} K., {Dalla} S., {Patsou} I., {Marsh} M.~S.,
  {Crosby} N.~B., {Malandraki} O., {Tsiropoula} G., 2015, Sol. Phys., 290, 841

\bibitem[{{Echer} {et~al}\mbox{.}(2008){Echer}, {Gonzalez}, {Tsurutani}, \&
  {Gonzalez}}]{Echer2008}
{Echer} E., {Gonzalez} W.~D., {Tsurutani} B.~T., {Gonzalez} A.~L.~C., 2008, J.
  Geophys. Res. (Space Physics), 113, 5221

\bibitem[{{Feynman} \& {Gabriel}(2000)}]{Feynman2000}
{Feynman} J., {Gabriel} S.~B., 2000, J. Geophys. Res., 105, 10543

\bibitem[{{Fry} {et~al}\mbox{.}(2003){Fry}, {Dryer}, {Smith}, {Sun}, {Deehr},
  \& {Akasofu}}]{Fry2003}
{Fry} C.~D., {Dryer} M., {Smith} Z., {Sun} W., {Deehr} C.~S., {Akasofu} S.-I.,
  2003, J. Geophys. Res. (Space Physics), 108, 1070

\bibitem[{{Gopalswamy} {et~al}\mbox{.}(2001){Gopalswamy}, {Lara}, {Yashiro},
  {Kaiser}, \& {Howard}}]{Gopalswamy2001}
{Gopalswamy} N., {Lara} A., {Yashiro} S., {Kaiser} M.~L., {Howard} R.~A., 2001,
  J. Geophys. Res., 106, 29207

\bibitem[{{Gopalswamy} {et~al}\mbox{.}(2009){Gopalswamy}, {M{\"a}kel{\"a}},
  {Xie}, {Akiyama}, \& {Yashiro}}]{Gopalswamy2009}
{Gopalswamy} N., {M{\"a}kel{\"a}} P., {Xie} H., {Akiyama} S., {Yashiro} S.,
  2009, J. Geophys. Res. (Space Physics), 114, 0

\bibitem[{{Gosling} {et~al}\mbox{.}(1990){Gosling}, {Bame}, {McComas}, \&
  {Phillips}}]{Gosling1990}
{Gosling} J.~T., {Bame} S.~J., {McComas} D.~J., {Phillips} J.~L., 1990,
  Geophys. Res. Lett., 17, 901

\bibitem[{{Gosling} {et~al}\mbox{.}(1991){Gosling}, {McComas}, {Phillips}, \&
  {Bame}}]{Gosling1991}
{Gosling} J.~T., {McComas} D.~J., {Phillips} J.~L., {Bame} S.~J., 1991, J.
  Geophys. Res., 96, 7831

\bibitem[{{Gosling} {et~al}\mbox{.}(1987){Gosling}, {Thomsen}, {Bame}, \&
  {Zwickl}}]{Gosling1987}
{Gosling} J.~T., {Thomsen} M.~F., {Bame} S.~J., {Zwickl} R.~D., 1987, J.
  Geophys. Res., 92, 12399

\bibitem[{{Gui} {et~al}\mbox{.}(2011){Gui}, {Shen}, {Wang}, {Ye}, {Liu},
  {Wang}, \& {Zhao}}]{Gui2011}
{Gui} B., {Shen} C., {Wang} Y., {Ye} P., {Liu} J., {Wang} S., {Zhao} X., 2011,
  Sol. Phys., 271, 111

\bibitem[{Gurney(1997)}]{Gurney1997}
Gurney K., 1997, {An Introduction to Neural Networks}. UCL Press, London, UK

\bibitem[{{Li} \& {Zhu}(2013)}]{Li2013}
{Li} R., {Zhu} J., 2013, Research in Astronomy and Astrophysics, 13, 1118

\bibitem[{{Lugaz} {et~al}\mbox{.}(2011){Lugaz}, {Downs}, {Shibata}, {Roussev},
  {Asai}, \& {Gombosi}}]{Lugaz2011}
{Lugaz} N., {Downs} C., {Shibata} K., {Roussev} I.~I., {Asai} A., {Gombosi}
  T.~I., 2011, ApJ, 738, 127

\bibitem[{{Lugaz} {et~al}\mbox{.}(2012){Lugaz}, {Farrugia}, {Davies},
  {M{\"o}stl}, {Davis}, {Roussev}, \& {Temmer}}]{Lugaz2012}
{Lugaz} N., {Farrugia} C.~J., {Davies} J.~A., {M{\"o}stl} C., {Davis} C.~J.,
  {Roussev} I.~I., {Temmer} M., 2012, ApJ, 759, 68

\bibitem[{{Manoharan}(2006)}]{Manoharan2006}
{Manoharan} P.~K., 2006, Sol. Phys., 235, 345

\bibitem[{{Mari{\v c}i{\'c}} {et~al}\mbox{.}(2014){Mari{\v c}i{\'c}}, {Vr{\v
  s}nak}, {Dumbovi{\'c}}, {{\v Z}ic}, {Ro{\v s}a}, {Hr{\v z}ina}, {Luli{\'c}},
  {Rom{\v s}tajn}, {Bu{\v s}i{\'c}}, {Salamon}, {Temmer}, {Rollett}, {Veronig},
  {Bostanjyan}, {Chilingarian}, {Mailyan}, {Arakelyan}, {Hovhannisyan}, \&
  {Muji{\'c}}}]{Maricic2014}
{Mari{\v c}i{\'c}} D. {et~al.}, 2014, Sol. Phys., 289, 351

\bibitem[{{Marubashi} {et~al}\mbox{.}(2015){Marubashi}, {Akiyama}, {Yashiro},
  {Gopalswamy}, {Cho}, \& {Park}}]{Marubashi2015}
{Marubashi} K., {Akiyama} S., {Yashiro} S., {Gopalswamy} N., {Cho} K.-S.,
  {Park} Y.-D., 2015, Sol. Phys., 290, 1371

\bibitem[{{Mays} {et~al}\mbox{.}(2015){Mays}, {Taktakishvili}, {Pulkkinen},
  {MacNeice}, {Rast{\"a}tter}, {Odstrcil}, {Jian}, {Richardson}, {LaSota},
  {Zheng}, \& {Kuznetsova}}]{Mays2015}
{Mays} M.~L. {et~al.}, 2015, Sol. Phys., 290, 1775

\bibitem[{{M{\"o}stl} {et~al}\mbox{.}(2014){M{\"o}stl}, {Amla}, {Hall},
  {Liewer}, {De Jong}, {Colaninno}, {Veronig}, {Rollett}, {Temmer}, {Peinhart},
  {Davies}, {Lugaz}, {Liu}, {Farrugia}, {Luhmann}, {Vr{\v s}nak}, {Harrison},
  \& {Galvin}}]{Mostl2014}
{M{\"o}stl} C. {et~al.}, 2014, ApJ, 787, 119

\bibitem[{{Odstrcil}, {Riley} \& {Zhao}(2004){Odstrcil}, {Riley}, \&
  {Zhao}}]{Odstrcil2004}
{Odstrcil} D., {Riley} P., {Zhao} X.~P., 2004, J. Geophys. Res., 109, 2116

\bibitem[{{Pr{\v s}a} {et~al}\mbox{.}(2008){Pr{\v s}a}, {Guinan}, {Devinney},
  {DeGeorge}, {Bradstreet}, {Giammarco}, {Alcock}, \& {Engle}}]{Prsa2008}
{Pr{\v s}a} A., {Guinan} E.~F., {Devinney} E.~J., {DeGeorge} M., {Bradstreet}
  D.~H., {Giammarco} J.~M., {Alcock} C.~R., {Engle} S.~G., 2008, ApJ, 687, 542

\bibitem[{{Richardson} \& {Cane}(2010)}]{Richardson2010}
{Richardson} I.~G., {Cane} H.~V., 2010, Sol. Phys., 264, 189

\bibitem[{{Richardson} \& {Cane}(2012)}]{Richardson2012}
{Richardson} I.~G., {Cane} H.~V., 2012, Journal of Space Weather and Space
  Climate, 2, A1

\bibitem[{{Richardson}, {Cliver} \& {Cane}(2001){Richardson}, {Cliver}, \&
  {Cane}}]{Richardson2001}
{Richardson} I.~G., {Cliver} E.~W., {Cane} H.~V., 2001, Geophys. Res. Lett, 28,
  2569

\bibitem[{{Rollett} {et~al}\mbox{.}(2014){Rollett}, {M{\"o}stl}, {Temmer},
  {Frahm}, {Davies}, {Veronig}, {Vr{\v s}nak}, {Amerstorfer}, {Farrugia}, {{\v
  Z}ic}, \& {Zhang}}]{Rollett2014}
{Rollett} T. {et~al.}, 2014, ApJL, 790, L6

\bibitem[{{Schrijver} \& {Mitchell}(2013)}]{Schrijver2013}
{Schrijver} C.~J., {Mitchell} S.~D., 2013, Journal of Space Weather and Space
  Climate, 3, A19

\bibitem[{{Schwenn} {et~al}\mbox{.}(2005){Schwenn}, {Dal Lago}, {Huttunen}, \&
  {Gonzalez}}]{Schwenn2005}
{Schwenn} R., {Dal Lago} A., {Huttunen} E., {Gonzalez} W.~D., 2005, Annales
  Geophysicae, 23, 1033

\bibitem[{{Shen} {et~al}\mbox{.}(2011){Shen}, {Wang}, {Gui}, {Ye}, \&
  {Wang}}]{Shen2011}
{Shen} C., {Wang} Y., {Gui} B., {Ye} P., {Wang} S., 2011, Sol. Phys., 269, 389

\bibitem[{{Taktakishvili} {et~al}\mbox{.}(2009){Taktakishvili}, {Kuznetsova},
  {MacNeice}, {Hesse}, {Rast{\"a}tter}, {Pulkkinen}, {Chulaki}, \&
  {Odstrcil}}]{Taktakishvili2009}
{Taktakishvili} A., {Kuznetsova} M., {MacNeice} P., {Hesse} M., {Rast{\"a}tter}
  L., {Pulkkinen} A., {Chulaki} A., {Odstrcil} D., 2009, Space Weather, 7, 3004

\bibitem[{{Temmer} {et~al}\mbox{.}(2011){Temmer}, {Rollett}, {M{\"o}stl},
  {Veronig}, {Vr{\v s}nak}, \& {Odstr{\v c}il}}]{Temmer2011}
{Temmer} M., {Rollett} T., {M{\"o}stl} C., {Veronig} A.~M., {Vr{\v s}nak} B.,
  {Odstr{\v c}il} D., 2011, ApJ, 743, 101

\bibitem[{{Temmer} {et~al}\mbox{.}(2012){Temmer}, {Vr{\v s}nak}, {Rollett},
  {Bein}, {de Koning}, {Liu}, {Bosman}, {Davies}, {M{\"o}stl}, {{\v Z}ic},
  {Veronig}, {Bothmer}, {Harrison}, {Nitta}, {Bisi}, {Flor}, {Eastwood},
  {Odstrcil}, \& {Forsyth}}]{Temmer2012}
{Temmer} M. {et~al.}, 2012, ApJ, 749, 57

\bibitem[{{Uwamahoro}, {McKinnell} \& {Habarulema}(2012){Uwamahoro},
  {McKinnell}, \& {Habarulema}}]{Uwamahoro2012}
{Uwamahoro} J., {McKinnell} L.~A., {Habarulema} J.~B., 2012, Annales
  Geophysicae, 30, 963

\bibitem[{{Valach} {et~al}\mbox{.}(2009){Valach}, {Revallo}, {Bochn{\'{\i}}{\v
  c}ek}, \& {Hejda}}]{Valach2009}
{Valach} F., {Revallo} M., {Bochn{\'{\i}}{\v c}ek} J., {Hejda} P., 2009, Space
  Weather, 7, 4004

\bibitem[{{Vr{\v s}nak} {et~al}\mbox{.}(2004){Vr{\v s}nak}, {Ru{\v z}djak},
  {Sudar}, \& {Gopalswamy}}]{Vrsnak2004}
{Vr{\v s}nak} B., {Ru{\v z}djak} D., {Sudar} D., {Gopalswamy} N., 2004, A\&A,
  423, 717

\bibitem[{{Vr{\v s}nak}, {Sudar} \& {Ru{\v z}djak}(2005){Vr{\v s}nak}, {Sudar},
  \& {Ru{\v z}djak}}]{Vrsnak2005}
{Vr{\v s}nak} B., {Sudar} D., {Ru{\v z}djak} D., 2005, A\&A, 435, 1149

\bibitem[{{Vr{\v s}nak} {et~al}\mbox{.}(2014){Vr{\v s}nak}, {Temmer}, {{\v
  Z}ic}, {Taktakishvili}, {Dumbovi{\'c}}, {M{\"o}stl}, {Veronig}, {Mays}, \&
  {Odstr{\v c}il}}]{Vrsnak2014}
{Vr{\v s}nak} B. {et~al.}, 2014, ApJS, 213, 21

\bibitem[{{Vr{\v s}nak} \& {{\v Z}ic}(2007)}]{Vrsnak2007}
{Vr{\v s}nak} B., {{\v Z}ic} T., 2007, A\&A, 472, 937

\bibitem[{{Vr{\v s}nak} {et~al}\mbox{.}(2013){Vr{\v s}nak}, {{\v Z}ic},
  {Vrbanec}, {Temmer}, {Rollett}, {M{\"o}stl}, {Veronig}, {{\v C}alogovi{\'c}},
  {Dumbovi{\'c}}, {Luli{\'c}}, {Moon}, \& {Shanmugaraju}}]{Vrsnak2013}
{Vr{\v s}nak} B. {et~al.}, 2013, Sol. Phys., 285, 295

\bibitem[{{Wang} {et~al}\mbox{.}(2004){Wang}, {Shen}, {Wang}, \&
  {Ye}}]{Wang2004}
{Wang} Y., {Shen} C., {Wang} S., {Ye} P., 2004, Sol. Phys., 222, 329

\bibitem[{{Wang} {et~al}\mbox{.}(2002){Wang}, {Ye}, {Wang}, {Zhou}, \&
  {Wang}}]{Wang2002}
{Wang} Y.~M., {Ye} P.~Z., {Wang} S., {Zhou} G.~P., {Wang} J.~X., 2002, J.
  Geophys. Res. (Space Physics), 107, 1340

\bibitem[{{Zhang} {et~al}\mbox{.}(2003){Zhang}, {Dere}, {Howard}, \&
  {Bothmer}}]{Zhang2003}
{Zhang} J., {Dere} K.~P., {Howard} R.~A., {Bothmer} V., 2003, ApJ, 582, 520

\bibitem[{{Zhang} {et~al}\mbox{.}(2007){Zhang}, {Richardson}, {Webb},
  {Gopalswamy}, {Huttunen}, {Kasper}, {Nitta}, {Poomvises}, {Thompson}, {Wu},
  {Yashiro}, \& {Zhukov}}]{Zhang2007}
{Zhang} J. {et~al.}, 2007, J. Geophys. Res. (Space Physics), 112, 10102

\bibitem[{{Zhao} \& {Dryer}(2014)}]{Zhao2014}
{Zhao} X., {Dryer} M., 2014, Space Weather, 12, 448

\end{thebibliography}

\end{document}